\documentclass[conference,compsoc]{IEEEtran}
\ifCLASSOPTIONcompsoc
  \usepackage[nocompress]{cite}
\else
  \usepackage{cite}
\fi
\ifCLASSINFOpdf
\else
\fi
\usepackage{amsmath}
\usepackage{graphicx}

\usepackage{array}
\usepackage{multirow}

\usepackage{booktabs}

\begin{document}
%
\title{Bounded Autonomy: A Trust-Anchored Defense Against Resource-Exhaustion Attacks on Tool-Augmented LLM Agents}

\author{\IEEEauthorblockN{
Ziyuan Chen\IEEEauthorrefmark{1},
Yueming Lyu\IEEEauthorrefmark{1},
Songze Li\IEEEauthorrefmark{2},
Huafeng Chen\IEEEauthorrefmark{1},
Caifeng Shan\IEEEauthorrefmark{1}
}
\IEEEauthorblockA{
\IEEEauthorrefmark{1}
School of Intelligence Science and Technology, Nanjing University\\
Suzhou, China\\
Email: ymlv@nju.edu.cn, cfshan@nju.edu.cn
}
\IEEEauthorblockA{
\IEEEauthorrefmark{2}
School of Cyber Science and Engineering, Southeast University\\
Nanjing, China\\
}}


%


\maketitle

\begin{abstract}
Tool-augmented Large Language Model (LLM) agents are vulnerable to resource-exhaustion attacks that amplify model-token consumption through persistent context amplification, large tool output amplification, and error-induced agentic loops. Despite growing attention to agent security and resource-abuse risks, a systematic defense framework specifically targeting these attacks remains lacking. Meanwhile, existing general-purpose agent defense frameworks, which address a broader range of security threats, often rely on prompt-level safeguards or coarse-grained resource limits and lack deterministic runtime mechanisms for controlling token consumption. They also provide limited support for balancing fault recovery against the risk of repeated resource-consuming execution. To address these limitations, we propose a trust-anchored closed-loop defense framework consisting of four phases: certificate-guided orchestration with deterministic handoffs, calibrated one-time node budgeting, Gateway-based runtime enforcement, and bounded recovery through a shared, non-resettable recovery pool. By accounting for both input and output tokens across all model invocations, our experiments demonstrate that the framework bounds cumulative token consumption, mitigates resource-amplification attacks, and preserves benign task utility under trusted execution assumptions.
\end{abstract}


%
\IEEEpeerreviewmaketitle

\section{Introduction}

Recently, tool-augmented Large Language Model (LLM) agents~\cite{yao2023react,schick2023toolformer} have emerged as a dominant paradigm for automating complex multi-step tasks across diverse domains, ranging from automated software development to enterprise workflow automation. By integrating LLM-driven autonomous planning with external execution environments, including APIs, databases, and web interfaces, these agents can dynamically retrieve external information, execute domain-specific tools, and iteratively refine their actions. To support continuous operation across heterogeneous tool ecosystems, modern agent architectures~\cite{wu2024autogen,hong2024metagpt} increasingly employ trusted orchestration components to coordinate task execution, maintain execution states, and manage access to external resources.

However, the autonomous and iterative nature of LLM agents also exposes them to resource-exhaustion attacks~\cite{317985}. By manipulating external information and execution feedback, adversaries can induce excessive model interactions, inflate context across successive requests, and prolong execution without meaningful progress~\cite{zhang2025breaking,li2026sponge,hou2026agents}. The resulting amplification can have significant economic and operational consequences: a recent study~\cite{zhou-etal-2026-beyond-max} reported up to 658$\times$ higher per-query costs, 100--560$\times$ higher energy consumption, and 35--74\% GPU KV-cache occupancy. Such attacks can impose unexpected API expenses on individual users and substantial unplanned costs and resource contention on enterprises. Unlike conventional attacks targeting data confidentiality or system integrity, resource-exhaustion attacks exploit legitimate agent execution mechanisms to amplify token consumption without necessarily preventing task completion. Consequently, an agent may accomplish its objective while consuming far more resources than necessary, leaving substantial costs hidden behind seemingly normal task progression.

\begin{figure}[!t]
    \centering
    \includegraphics[width=1.0\linewidth]{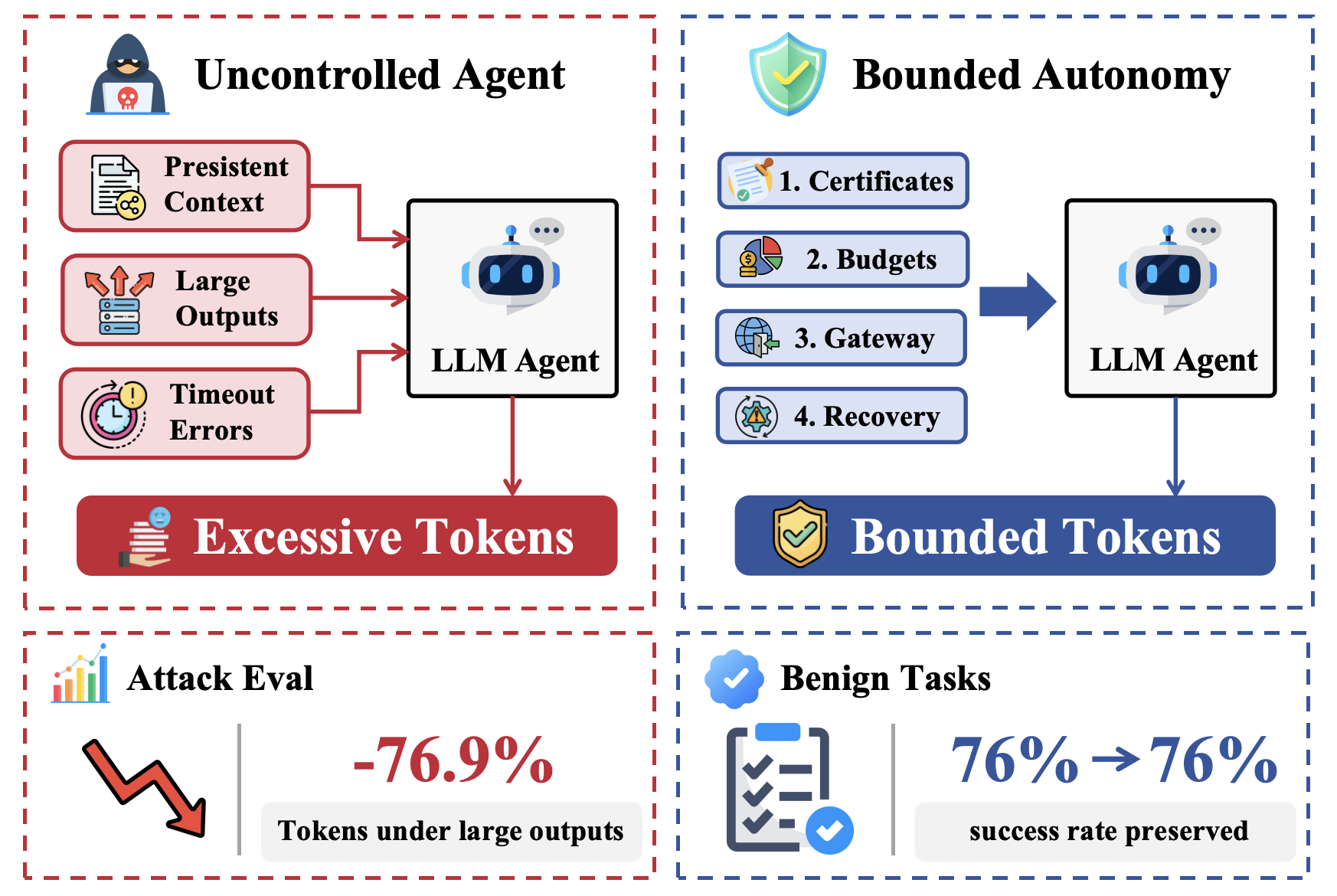}
    \caption{Motivation and key idea of Bounded Autonomy. }
    \label{fig:0}
\end{figure}

Despite growing attention to agent security and resource-abuse risks, a systematic defense framework specifically targeting resource-exhaustion attacks remains lacking. Meanwhile, existing agent-security research~\cite{debenedetti2024agentdojo,zhan2024injecagent,ruan2024identifying,zhang2025agent,debenedetti2026defeating,wang2026agentspec,jia2025task} has primarily focused on prompt-injection robustness, harmful tool use, and risk assessment, rather than providing comprehensive resource governance throughout the execution lifecycle. Their reliance on prompt-level safeguards or coarse-grained resource limits cannot establish deterministic control over actual token consumption. Furthermore, existing failure-handling strategies often face a trade-off between allowing repeated retries, which may provide adversaries with additional opportunities for resource exhaustion, and terminating execution immediately, which can discard valid intermediate progress and degrade fault tolerance~\cite{zhang2026persistent}. These limitations motivate a unified defense framework that can regulate task progression and resource consumption while retaining the ability to recover from execution interruptions.

To address these challenges, we propose a trust-anchored closed-loop defense framework that separates agent autonomy from system-level execution governance. By keeping the agent's internal planning flexible while subjecting task advancement, execution-budget allocation, and recovery admission to independent Host-side verification and enforcement, our framework preserves task-solving autonomy without relying on the agent to regulate its own progress or resource consumption. The framework operates through four interconnected phases. First, task execution is organized around a Host-governed structure in which progress must be independently verified before subsequent work can proceed. Second, once a subtask becomes eligible for execution, it receives a one-time token authorization derived from a calibrated reference demand and its initial request quote. This prevents repeated attempts or changes in execution identity from generating additional ordinary resources. Third, all model and tool interactions are mediated by a centralized enforcement boundary, where resource consumption is accounted for and constrained throughout execution. Finally, when execution is interrupted, the framework suspends ordinary resource allocation and permits recovery only within a separate, finite resource pool. Importantly, these phases are not independent safeguards: verified task progress determines when resources can be allocated, allocated budgets constrain subsequent execution, runtime enforcement governs every resource-consuming interaction, and bounded recovery preserves verified progress without resetting existing resource limits. Together, they establish a continuous governance loop that maintains agent autonomy while preventing uncontrolled resource expenditure.

In summary, our key contributions are as follows:

\textit{(i)} \textbf{A Trust-Anchored Closed-Loop Defense Framework: } We propose a unified defense framework for mitigating resource-exhaustion attacks against tool-augmented LLM agents. By coordinating trusted task progression, upfront resource allocation, runtime enforcement, and bounded recovery, the framework controls resource consumption without unnecessarily restricting normal agent autonomy.

\textit{(ii)} \textbf{Deterministic Token Resource Governance: } We develop a task-level calibrated reference demand estimator and a node-specific one-time token authorization mechanism, together with centralized runtime enforcement to regulate token allocation and consumption throughout task execution. The framework further incorporates a shared, non-resettable recovery mechanism to prevent repeated execution attempts from bypassing resource limits.

\textit{(iii)} \textbf{Empirical Validation of Security and Utility: } We evaluate our framework across benign workflows and adversarial attack scenarios. By accounting for both input and output tokens across all model invocations, including auxiliary and recovery-related calls, we demonstrate its effectiveness in bounding cumulative token consumption, mitigating resource-amplification attacks, and preserving benign task execution utility.

\section{Related Work}

\subsection{LLM Agent Security}

Agent security extends beyond indirect prompt injection to unsafe tool use, privacy leakage, harmful requests, and manipulation of agent memory or planning. AgentDojo~\cite{debenedetti2024agentdojo} and InjecAgent~\cite{zhan2024injecagent} study indirect prompt injection.  ToolSword~\cite{ye2024toolsword} and ToolEmu~\cite{ruan2024identifying} examine broader tool-use risks. R-Judge~\cite{yuan2024r} evaluates behavioral risk awareness, PrivacyLens~\cite{shao2024privacylens} studies privacy leakage, and AgentHarm~\cite{andriushchenko2025agentharm} assesses harmful multi-step requests. Agent Security Bench (ASB)~\cite{zhang2025agent} further covers attacks such as memory poisoning and plan-of-thought backdoors. While these works characterize unsafe agent behaviors and evaluate robustness, resource exhaustion arising from compromised execution remains less explored.

\subsection{Resource Exhaustion Attacks and Defenses}

Recent studies have begun to characterize resource-oriented threats in LLM agents. Persistent Billable State~\cite{zhang2026persistent} investigates recurring input costs from retained tool outputs and proposes host-side safeguards against context-driven cost amplification. Sponge Tool Attack (STA)~\cite{li2026sponge} induces verbose reasoning through adversarial prompt rewriting. Breaking Agents~\cite{zhang2025breaking} amplifies failures through repetitive or irrelevant actions, and AgentDoS~\cite{317985} uses directed grey-box fuzzing to uncover resource-exhaustion vulnerabilities. While existing defenses address specific resource-amplification mechanisms, our work focuses on lifecycle-wide resource governance by integrating verified task progress, one-time node budgeting, runtime accounting, and bounded recovery.

\subsection{Agent Execution Control}

Prior work has explored workflow orchestration and runtime enforcement to structure agent execution and constrain unsafe behavior. ReAct~\cite{yao2023react} interleaves reasoning and actions; StateFlow~\cite{wu2024stateflow} adopts state-driven workflows; MetaGPT~\cite{hong2024metagpt} incorporates standardized procedures into multi-agent collaboration; and AFlow~\cite{zhang2025aflow} automates the generation and optimization of agentic workflows. For runtime enforcement, NeMo Guardrails~\cite{rebedea-etal-2023-nemo} provides programmable conversational constraints, GuardAgent~\cite{xiang2025guardagent} generates executable checks for agent actions, RvLLM~\cite{zhang2025rvllm} verifies outputs against domain-specific specifications, and AgentSpec~\cite{wang2026agentspec} supports customizable runtime constraints. While these approaches improve workflow organization or enforce behavioral policies, they do not jointly couple Host-authenticated task-progress verification with one-time node-level token authorization, continuous resource accounting, and bounded recovery.

\section{Threat Model}

\subsection{System Model}

\begin{figure}[!t]
    \centering
    \includegraphics[width=1.0\linewidth]{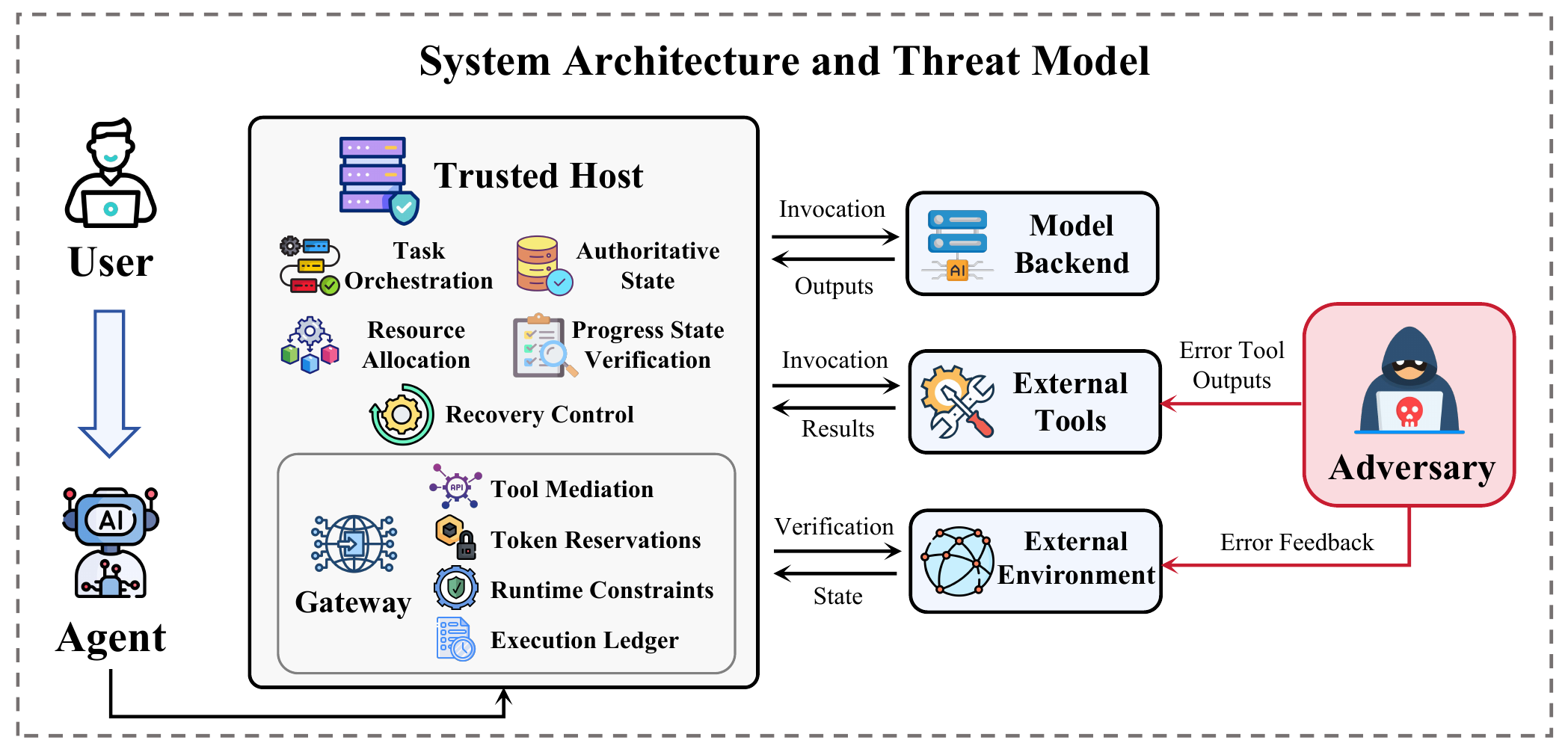}
    \caption{System architecture and threat model of the tool-augmented LLM agent system. The trusted Host incorporates a Gateway that mediates all model and tool invocations, while the adversary manipulates external tool outputs, content, and error feedback to induce excessive resource consumption.}
    \label{fig:1}
\end{figure}

\subsubsection{System Architecture}

Following prior formulations of tool-agent-user interactions and stateful execution environments~\cite{26dddae50d5a4610bdaced3e9fdb1c89,christodorescu2026agent,lu2025toolsandbox}, as well as agent-computer interfaces for interacting with external execution environments~\cite{yang2024sweagent,wang2025openhands}, we model the system as four entities, as illustrated in Figure~\ref{fig:1}:

\textit{(i) User. }The user initiates an execution by submitting a high-level task request and may provide additional information or clarifications during execution.

\textit{(ii) Agent. }An LLM-based autonomous system that plans and executes tasks through iterative model and tool interactions.

\textit{(iii) Host. }A trusted orchestration component that manages task execution, maintains authoritative task states, allocates resources, verifies progress, and controls recovery.

\textit{(iv) External Tools and Environment. }Tools provide information retrieval and action interfaces, while the environment maintains the underlying application state and reflects action effects.

The Host retains authoritative control over task orchestration and security policies. Its logically distinct Gateway mediates all model and tool invocations and enforces these policies at the execution boundary.

\subsubsection{Agent Execution Process}

Following prior work on structured agent workflows and stateful tool execution~\cite{yue2026static,li2026beyond,wang2026lean4agent}, we model task execution as follows:

\textit{(i) Task Decomposition. }Given a user task, the Host compiles a fixed authorization DAG $G=(V,E)$ from supported task contracts. Each node represents an authorized business subtask or obligation, specifying its allowed tools, input/output contract, Host completion predicate, and node-level token budget account. Edges define dependencies between subtasks. The DAG remains frozen throughout execution: the agent may plan and re-plan within a node but cannot modify the authorization DAG or create additional budget-bearing nodes.

\textit{(ii) Node Execution. }The agent selects a \textsc{Ready} node and iteratively invokes LLMs and external tools to accomplish its objective, incorporating returned results into subsequent interactions. All invocations pass through the Gateway and comply with the node's resource authorization and execution constraints.

\textit{(iii) Node Completion Verification. }Upon node termination, the Host validates its outcome against the corresponding completion conditions in the external environment. A node is completed only after its required business conditions are independently verified and a valid Host-authenticated completion certificate is issued.

\textit{(iv) Task Progression. }Host-authenticated completion certificates enable dependent nodes in the authorization DAG to become executable. Execution continues until all required subtasks are completed or the task terminates.

\subsubsection{Resource Consumption Model}

The primary resource considered in this work is model-token consumption, including input and output tokens across all LLM invocations. This covers normal node execution, auxiliary control operations, and recovery-related model calls.

For each model invocation $a$, let $q_a$ denote its complete pre-dispatch token quote and $c_a$ its actual settled token consumption, including both input and output tokens. We assume a trusted execution contract that accounts for the serialized request before dispatch and enforces the configured maximum output-token allowance. Thus, every valid invocation satisfies:

\begin{equation}
0 \leq c_a \leq q_a.
\end{equation}

This upper bound is enforced by trusted execution infrastructure, not by the agent's self-reported usage. Any violation constitutes an infrastructure failure outside our normal threat model.

Native tool execution does not necessarily incur direct model-token consumption, but its invocation count, execution duration, and returned payload size remain subject to independent runtime constraints. Tool outputs contribute to model-token consumption when incorporated into subsequent model input contexts.

Accordingly, the system distinguishes the token-budget ledger from other runtime constraints. The former tracks cumulative input and output tokens across model invocations; the latter includes model and tool invocation limits, per-call timeouts, tool-result size limits, and global execution deadlines. The Gateway and trusted execution infrastructure enforce these constraints independently.

For evaluation, cumulative model-token consumption is measured across all model invocations:

\begin{equation}
C_{\mathrm{total}} =
\sum_{a\in\mathcal A_{\mathrm{model}}} c_a,
\end{equation}

where $\mathcal A_{\mathrm{model}}$ covers normal execution, auxiliary control operations, and recovery-related calls. No invocation is excluded based on whether it occurs inside ordinary node execution.

\subsection{Adversary Model}

We consider an external adversary capable of manipulating information accessible to the agent through external tools and data sources~\cite{greshake2023not,
debenedetti2024agentdojo,zhan2024injecagent}. Rather than directly compromising the agent or Host, the adversary exploits autonomous execution and interaction mechanisms to induce excessive model-token consumption.

\subsubsection{Adversary Capabilities and Assumptions}

The adversary can manipulate tool outputs, external documents, database records, error messages, and intermediate results consumed by the agent~\cite{zhang2025agent}. By introducing redundant, oversized, or misleading content, it can induce repeated model and tool invocations, unnecessary context processing, and unproductive execution loops. However, the adversary cannot modify the Host's resource ledger, budget policies, authorization DAG, completion predicates, model and tokenizer configuration, Host authentication state, or recovery pool. Nor can it bypass the Gateway or directly alter trusted execution infrastructure.

\subsubsection{Attack Strategies}

Prior studies have revealed diverse resource-exhaustion behaviors in tool-augmented agents. This work focuses on three representative attack strategies that capture distinct and recurring resource-amplification mechanisms across the agent execution lifecycle. These strategies represent the scope of our threat model rather than an exhaustive taxonomy of all possible resource-exhaustion attacks.

\textit{(i) Persistent Context Amplification. }The adversary injects redundant or misleading content into legitimate tool outputs, causing it to persist across subsequent model contexts and repeatedly increase input-token consumption~\cite{zhang2026persistent}.

\textit{(ii) Large Tool Output Amplification. }The adversary introduces oversized, repetitive, or structurally complex tool outputs, increasing subsequent input-token consumption or inducing verbose responses and unnecessary tool invocations~\cite{zhou-etal-2026-beyond-max,lee2026overthinking}.

\textit{(iii) Error-Induced Agentic Loops. }The adversary provides repetitive errors, timeouts, or invalid feedback to induce repeated retries, parameter modifications, or replanning without meaningful task progress~\cite{zhang2025breaking,hou2026agents,317985}.

\subsubsection{Adversary Goals}

The adversary aims to inflate model-token consumption beyond legitimate task requirements, imposing unnecessary API costs on users or service operators and potentially exhausting execution budgets or interrupting tasks. It seeks to waste paid resources and disrupt service through costly interactions that produce little or no meaningful business progress. An attack can succeed even if the agent eventually completes the original task, since excess consumption and the resulting financial costs are themselves harmful outcomes.

\subsection{Defender Model}

\subsubsection{Defender Authority and Trust Assumptions}

We consider a managed agent deployment, such as an enterprise workflow or customer-service platform, where the service operator controls task orchestration, model API access, and registered tool integrations. The following assumptions define the trusted execution boundary:

\textit{(i) Host Integrity and Authority.} The Host runs as an operator-controlled orchestration service. Its internal state, including the authorization DAG, completion predicates, resource ledger, execution policies, and recovery budget, is maintained outside the agent's control. The Host can authorize, schedule, and suspend node execution.

\textit{(ii) Gateway Enforcement.} Model and tool requests are routed through a centralized API Gateway and tool broker. Backend credentials and direct access permissions are controlled by the platform, preventing the agent from bypassing the Gateway. The Gateway enforces Host-configured token reservations and runtime constraints, accounts for resource consumption, and controls result delivery.

\textit{(iii) Execution Infrastructure.} Supported model backends provide model-compatible input-token accounting, enforce maximum output-token limits, and expose usage records for verifying actual consumption. Tool adapters enforce configured timeouts and result-size limits. These mechanisms provide verifiable execution records for resource accounting.

\textit{(iv) Business-State Verification.} For supported tasks, the underlying business application exposes authoritative state through trusted APIs or database interfaces (e.g., order records or transaction status). The Host uses these interfaces to independently verify task completion, rather than treating agent-generated claims or tool responses as authoritative evidence.

We assume that the Host, Gateway, execution infrastructure, and verification interfaces remain uncompromised. Host compromise, Gateway bypass, violations of the trusted token-metering contract, and unobservable external side effects are therefore outside our threat model.

\begin{figure*}[!t]
    \centering
    \includegraphics[width=0.85\linewidth]{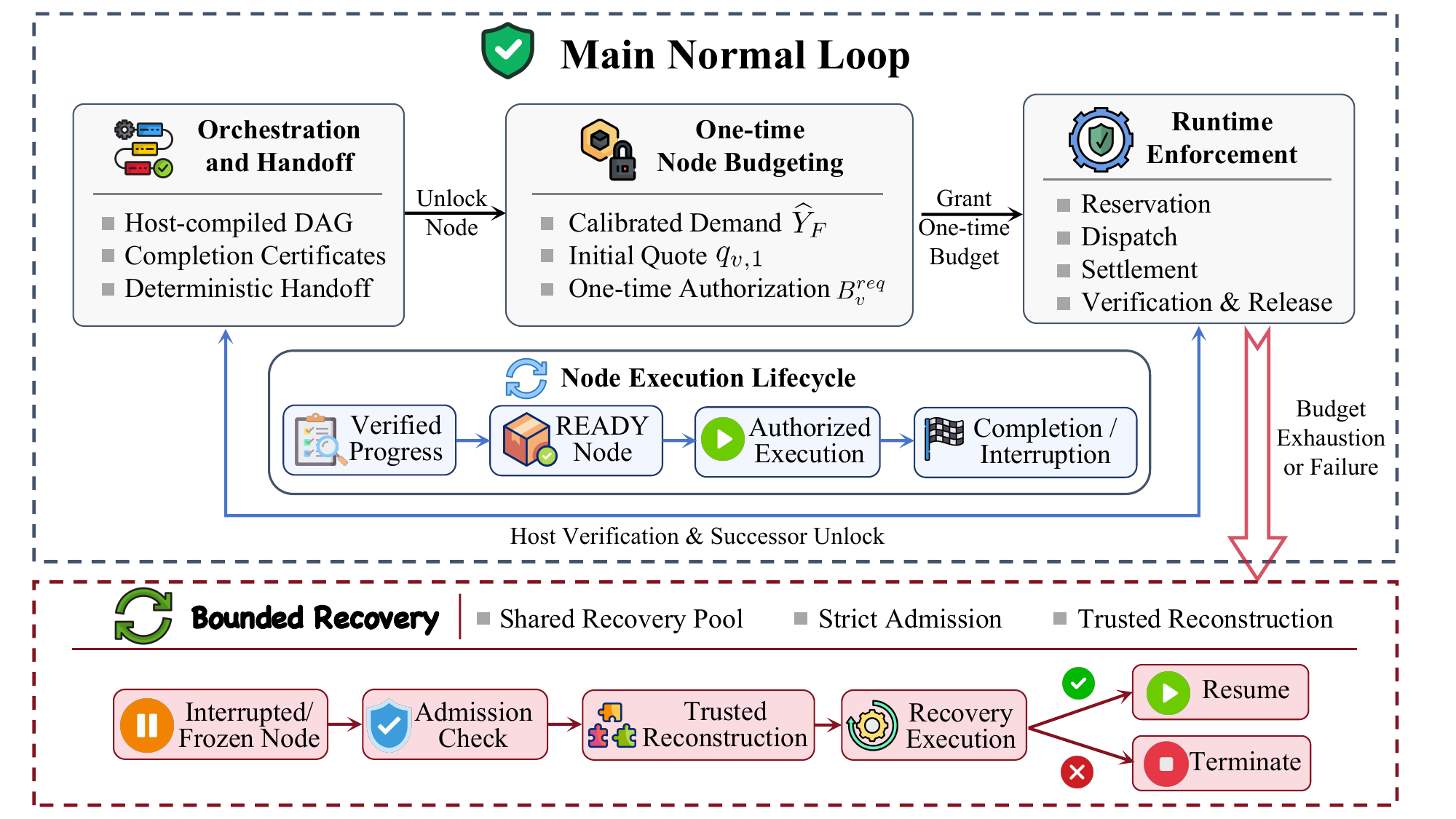}
    \caption{Overview of the proposed trust-anchored closed-loop defense framework. The main execution loop comprises certificate-guided orchestration, calibrated one-time node budgeting, and Gateway-based runtime enforcement. Budget exhaustion, threshold-qualified anomalies, or eligible transient failures may trigger a bounded recovery process governed by strict admission control, trusted state reconstruction, and a shared, non-resettable recovery pool.}
    \label{fig:2}
\end{figure*}

\subsubsection{Defender Objectives}

The defender aims to mitigate resource-exhaustion attacks while preserving normal agent utility. We consider four objectives:

\textit{(i) }\textbf{Normal Task Utility Preservation. }Maintain benign task success rates while minimizing unnecessary execution overhead and defense-induced false interruptions.

\textit{(ii) }\textbf{Bounded Model-Token Consumption. }Bound cumulative input and output token consumption across all model invocations, preventing unbounded resource expenditure caused by adversarial inputs or execution behaviors.

\textit{(iii) }\textbf{Timely Interruption and Damage Containment. }Detect abnormal execution behavior and promptly suspend ordinary execution to limit further token consumption and uncontrolled resource expenditure.

\textit{(iv) }\textbf{Bounded Recovery of Interrupted Nodes. }Following budget exhaustion, detected execution anomalies, or eligible transient operational failures, attempt to resume interrupted nodes within the original task using a finite recovery budget and bounded attempts, while preserving verified progress. Recovery is best-effort and is not guaranteed under arbitrary attack intensity or execution conditions.


\section{Method}

\subsection{Overview}

To mitigate resource-exhaustion attacks while preserving task utility, BA-Agent separates agent autonomy from system-level resource governance through four integrated mechanisms.

\textit{(i) Certificate-Guided Orchestration and Handoff.} A Host-compiled authorization DAG and Host-authenticated completion certificates govern task progression, while deterministic, bounded handoffs limit cross-node context accumulation.

\textit{(ii) Calibrated One-Time Node Budgeting.} The Host derives each node's one-time token authorization from a frozen, benign-data-calibrated reference demand and its initial request quote, preventing budget renewal through retries or replanning.

\textit{(iii) Gateway-Based Runtime Enforcement.} A centralized Gateway mediates all model and tool invocations through reservation, dispatch, settlement, and verification and release, enforcing global token accounting and runtime constraints.

\textit{(iv) Bounded Recovery.} A shared, non-resettable recovery pool with strict admission conditions and a run-level attempt limit enables bounded, best-effort recovery from execution interruptions.

\subsection{Certificate-Guided Orchestration and Handoff}
\label{sec:orchestration}

\subsubsection{Host-Compiled Authorization DAG}

Before execution, the Host loads and freezes an authorization DAG $G=(V,E)$ from supported task contracts. Each node $v\in V$ specifies \textit{(i)} a public business objective, \textit{(ii)} its predecessors, \textit{(iii)} allowed tools, \textit{(iv)} input/output contracts, \textit{(v)} a private Host completion predicate, \textit{(vi)} a required state version, \textit{(vii)} node-level guardrails, and \textit{(viii)} a token budget account.

The agent receives only the public graph view needed for its current execution frontier, without access to private completion predicates or their truth values. The DAG remains immutable: the agent cannot modify its nodes, dependencies, or completion predicates, or create additional budget-bearing nodes. This restriction anchors the one-time authorization mechanism in Section~\ref{sec:budget}.

\subsubsection{Node States and Readiness}

Nodes transition through the semantic states in Table~\ref{tab:node-states}; database enumerations may map to these states separately. Let $\mathcal C_t$ denote all nodes completed through valid, non-replayed Host-authenticated certificates issued by time $t$. These certificates remain valid throughout the run despite subsequent legitimate business-state changes, making $\mathcal C_t$ a cumulative set rather than one filtered by the current state version.

A node $v$ is executable if and only if its dependency condition holds over $\mathcal C_t$:

\begin{equation}
\mathrm{READY}(v,t)\iff\phi_v(\mathcal C_t)=1,
\label{eq:ready}
\end{equation}

where $\phi_v$ encodes the dependency requirements. AND-predecessors require all branches to hold, while OR-predecessors accept only a verified legal branch. Thus, readiness depends exclusively on Host-certified completions, not agent- or tool-reported progress.

\begin{table}[t]
\centering
\caption{Semantic node states. Database enumerations may map onto these states separately.}
\label{tab:node-states}
\begin{tabular}{@{}c p{0.72\linewidth}@{}}
\toprule
\textbf{State} & \textbf{Meaning and constraint} \\
\midrule
\textsc{Locked} & Predecessor condition unmet; cannot be selected and holds no ordinary budget. \\

\textsc{Ready} & Dependencies satisfied by $\mathcal C_t$; eligible for one-time ordinary token authorization. \\

\textsc{Executing} & Selected node may invoke models and tools through the Gateway under its granted budget. \\

\textsc{Verified} & Host has verified business state, side effects, and execution receipts, and issued certificate $C_v$. \\

\textsc{Frozen} & Normal scheduling suspended due to budget exhaustion, a threshold-qualified anomaly, transient operational failure, or untrusted state; in-flight liabilities are preserved. \\

\textsc{Recovering} & A bounded recovery attempt is active; ordinary scheduling remains closed and only the shared recovery pool is available. \\

\textsc{Terminated} & Node has finally ended (success or failure); it cannot receive another ordinary budget or be renamed to bypass the ledger. \\
\bottomrule
\end{tabular}
\end{table}

\subsubsection{Host-Authenticated Completion Certificates}

At a completion-verification boundary, the Host evaluates the node's private completion predicate against the external business state, allowed side effects, relevant state version, run identity, event prefix, resource settlement, and public output contract. Upon success, it issues:

\begin{equation}
C_v=\operatorname{Auth}_{\mathrm{Host}}\bigl(\mathcal I,\ v,\ \tau_v,\ h_G,\ e_v,\ R_v,\ \pi_v\bigr),
\label{eq:cert}
\end{equation}

which binds the verified completion fact to the run and business-state version observed at verification. Here, $\operatorname{Auth}_{\mathrm{Host}}$ denotes an abstract Host authentication operation; no specific cryptographic primitive is prescribed.

Table~\ref{tab:cert-symbols} summarizes the certificate fields. The certificate is bound to its run, node, and frozen authorization DAG, with integrity depending on its recorded verification state, event prefix, and execution receipts. Importantly, $\tau_v$ records the state version at verification; subsequent legitimate state changes do not invalidate this historical completion record. Certificates are accepted only within their original run and graph instance and must pass integrity and replay checks.

Certification is a deterministic Host procedure over trusted observations, execution receipts, and the private completion predicate, requiring no additional model call. Model-generated answers, ``completed'' strings in tool outputs, and self-reported progress cannot independently produce certificates. The Host may verify multiple pre-declared facts from one valid observation, but resource authorization remains independently deduplicated per node.

\begin{table}[t]
\centering
\caption{Symbols bound by the Host-authenticated completion certificate $C_v$.}
\label{tab:cert-symbols}
\begin{tabular}{@{}c p{0.8\linewidth}@{}}
\toprule
\textbf{Symbol} & \textbf{Meaning and binding} \\
\midrule
$\mathcal I$ & Immutable run identity. \\

$v$ & Certified logical DAG node. \\

$\tau_v$ & Business-state version at verification; records the verification context without requiring the current version to remain unchanged. \\

$h_G$ & Hash of the frozen Host-compiled authorization DAG, binding the certificate to the authorized graph. \\

$e_v$ & Event prefix belonging to this run and node execution. \\

$R_v$ & Binding to the node's settled Gateway receipts. \\

$\pi_v$ & Private Host completion predicate outcome. \\
\bottomrule
\end{tabular}
\end{table}

\subsubsection{Deterministic Business Handoff}

At a normal node boundary, the Host replaces the completed node's dialogue history with a deterministic handoff:

\begin{equation}
\begin{aligned}
\mathcal H_v &=
\operatorname{Project}_{v}\bigl(
\mathcal E_t^{\mathrm{delivered,valid}},
\mathcal O_v^{\mathrm{public}}\bigr),\\
&\qquad
\operatorname{Tok}_{\mathcal M}(\mathcal H_v)\leq L_{\mathrm{handoff}},
\end{aligned}
\label{eq:handoff}
\end{equation}

where $\mathcal E_t^{\mathrm{delivered,valid}}$ contains public facts actually delivered and still valid for the handoff, and $\mathcal O_v^{\mathrm{public}}$ denotes remaining public obligations at the current task frontier. Structured-field rules generate the handoff without an additional model summarization call.

The full transcript remains in the audit ledger but is not re-fed into subsequent model turns; the active node retains only messages needed for its execution. If required information exceeds the registered handoff length, the Host must reject the handoff or enter a defined exception path rather than silently truncate permissions, target objects, or unfinished obligations.

\subsubsection{Local Completion versus Final Success}

A completion certificate enables task progression and successor resource authorization but does not establish end-to-end task success. It records satisfaction of a node's completion predicate at its verification boundary, regardless of subsequent legitimate business-state changes. Final success is determined by the original task scorer on the actual final state, not by a handoff, recovery ticket, or agent closing answer, and is evaluated separately in Section~\ref{sec:experiments}.

\subsection{Calibrated One-Time Node Budgeting}
\label{sec:budget}

A completion certificate makes a node \textsc{Ready} but does not automatically grant resources. The Host computes a bounded token authorization using a fixed statistical reference demand estimator $F$, fitted on benign trajectories and calibrated at the task level. At runtime, its calibrated demand is combined with the node's initial request quote. Each logical node receives ordinary authorization at most once; retries, restarts, identity changes, and replanning cannot renew it. The Gateway enforces the grant through the ledger in Section~\ref{sec:gateway}.

\subsubsection{Reference Demand Formulation}

A node's authorized budget represents the minimum initial token credit needed to replay a complete, verified ordinary execution trajectory under serial dispatch. For node $v$ with call sequence $\{a_{v,1},\ldots,a_{v,m_v}\}$, let $q_{v,j}$ and $c_{v,j}$ denote the complete pre-dispatch quote and actual settled consumption of invocation $j$.

The complete request quote is

\begin{equation}
q_{v,j}=I_{v,j}^{\mathrm{quote}}+O_{\max},
\label{eq:request-quote}
\end{equation}

where

\begin{equation}
I_{v,j}^{\mathrm{quote}}=
\operatorname{Tok}_{\mathcal M}
\bigl(\operatorname{Serialize}(a_{v,j})\bigr)
+\delta_{\mathrm{meter}},
\label{eq:input-quote}
\end{equation}

and $\delta_{\mathrm{meter}}\geq 0$ is a frozen metering margin. Here, $\mathcal M$ specifies the fixed model, tokenizer, chat template, and tool-serialization contract. The serialized request includes all dispatched content, including system prompts, applicable tool schemas, public handoffs, and retained conversation history. The execution infrastructure enforces $O_{\max}$. Native tool calls may have zero direct model-token quotes but remain subject to independent invocation-count, execution-time, result-size, and side-effect constraints.

Let $I_{v,j}^{\mathrm{actual}}$ and $O_{v,j}^{\mathrm{actual}}$ denote actual input and output token consumption. Settled consumption is

\begin{equation}
c_{v,j}=I_{v,j}^{\mathrm{actual}}+O_{v,j}^{\mathrm{actual}}.
\label{eq:settled-consumption}
\end{equation}

Under the trusted execution contract in the Threat Model,

\begin{equation}
0\leq c_{v,j}\leq q_{v,j}.
\label{eq:quote-bound}
\end{equation}

For nodes requiring no model invocation, define $Y_v=0$. Otherwise, the required initial token credit is

\begin{equation}
Y_v=
\max\left\{
\sum_{j=1}^{m_v}c_{v,j},
\max_{1\leq j\leq m_v}
\left(
\sum_{\ell<j}c_{v,\ell}+q_{v,j}
\right)
\right\}.
\label{eq:demand}
\end{equation}

The first term captures total realized consumption; the second captures the maximum transient credit required before an invocation, including preceding settled consumption and the next request's full reservation. Only complete, verified ordinary trajectories provide demand observations. Interrupted, incomplete, or unverified trajectories are treated as lower-bound or unknown observations, not complete demand labels.

\subsubsection{Fixed Reference Estimation and Task-Level Calibration}

We construct a fixed reference estimator from benign trajectories using task-level demand aggregation and upper-tail calibration. Each training task is one statistical observation unit, with all its nodes kept together during fitting and calibration.

Let $\mathcal T$ denote the training tasks and $\mathcal D_t^{\mathrm{fit}}$ the eligible complete node observations for task $t$. Its aggregate demand is

\begin{equation}
Z_t=\max_{i\in\mathcal D_t^{\mathrm{fit}}}Y_i.
\label{eq:task-demand}
\end{equation}

Given reference level $\tau$, the baseline is the empirical quantile of task-level demands:

\begin{equation}
c_{\mathcal D}
=
\inf\left\{
z:
\frac{1}{|\mathcal T|}
\sum_{t\in\mathcal T}
\mathbf 1[Z_t\leq z]\geq\tau
\right\}.
\label{eq:weighted-quantile}
\end{equation}

For upper-tail calibration, training tasks are partitioned into $J$ folds, keeping all nodes of each task in the same fold. For task $t$ in fold $k(t)$, the baseline $c^{(-k(t))}$ is fitted on the other folds. Given positive scale parameter $s_0$, its out-of-fold score is

\begin{equation}
r_t=
\max_{i\in\mathcal D_t^{\mathrm{req}}}
\frac{Y_i-c^{(-k(t))}}{s_0},
\label{eq:task-residual}
\end{equation}

where $\mathcal D_t^{\mathrm{req}}$ contains all required node boundaries subject to demand prediction. Incomplete tasks, unmet original business-completion conditions, unknown required demand labels, or missing tool support receive $+\infty$. Fully verified tasks requiring no demand prediction receive zero scores; finite residuals retain their original signs.

Let $n=|\mathcal T|$ be the number of calibration tasks and $\alpha$ the upper-tail miscoverage parameter. Sorting scores as $r_{(1)}\leq\cdots\leq r_{(n)}$, we define

\begin{equation}
\ell=\left\lceil(n+1)(1-\alpha)\right\rceil,
\qquad
\widehat{\kappa}=r_{(\ell)}.
\label{eq:calibration-index}
\end{equation}

Calibration is unavailable if $\ell>n$ or the selected order statistic is non-finite. The baseline is then refitted on all eligible training tasks to obtain $c_{\mathrm{all}}$. The calibrated reference demand is

\begin{equation}
\widehat{Y}_F=
\left\lceil
\max\left\{0,c_{\mathrm{all}}+s_0\widehat{\kappa}\right\}
\right\rceil.
\label{eq:calibrated-demand}
\end{equation}

This procedure yields a fixed reference demand for the registered execution configuration. The upper-tail adjustment is empirical and does not provide a distribution-free finite-sample coverage guarantee. The estimator is fitted and calibrated on training data; hyperparameters and runtime configurations are selected on validation data and frozen before test evaluation. Further details are provided in the Appendix.

\subsubsection{Runtime Authorization and Safety Constraints}

At a node's first legal activation, the Host combines $\widehat{Y}_F$ with its initial request quote $q_{v,1}$ and applies a selected margin $\epsilon\geq 0$:

\begin{equation}
B_v^{\mathrm{req}}
=
\left\lceil
(1+\epsilon)\max\left\{\widehat{Y}_F,q_{v,1}\right\}
\right\rceil.
\label{eq:request}
\end{equation}

The Host grants this authorization atomically only if the node is \textsc{Ready}, its run and logical node identities are valid, its execution configuration belongs to the verified support set, and the estimator is valid for that configuration. It must also satisfy

\begin{equation}
B_v^{\mathrm{req}}
\leq
\min\left\{
\overline{B}_v,
E_{\mathrm{free}},
H-S-P
\right\},
\label{eq:grant}
\end{equation}

where $\overline{B}_v$ is the frozen node policy cap, $E_{\mathrm{free}}$ the unallocated ordinary token credit, $H$ the global token cap, and $S$ and $P$ the settled consumption and outstanding in-flight liabilities. The global cap is additionally enforced by the conservation invariant in Section~\ref{sec:gateway}.

If any condition fails, the Host records the rejection and binding constraint and suspends ordinary execution. It cannot silently truncate the requested budget, reduce the output allowance to admit an otherwise unaffordable request, or supplement ordinary authorization with recovery credits.

Upon authorization, the grant moves from the unallocated ordinary pool into the node's dedicated token account. Unused credits return to the ordinary pool only after Host verification or formal recovery settlement.

Ordinary grants are permanently persisted and deduplicated by logical node identity. Retries, restarts, context reconstruction, agent identity changes, and node renaming cannot create additional grants. The fitted baseline, calibration factor, support set, hyperparameters, execution margin, and runtime limits are frozen into run identity $\mathcal I$ before deployment.

\subsection{Gateway-Based Runtime Enforcement}
\label{sec:gateway}

All model and tool invocations must pass through the centralized Gateway within the trusted Host, including normal requests, tool calls, recovery planning, state reads, user-simulation turns, and auxiliary control calls. Every model invocation must be attributed to an authorized ordinary node account or the shared recovery pool before dispatch. The Gateway enforces Host-compiled budgets rather than independently estimating demand, preventing model-token consumption from bypassing the central ledger.

The Gateway accounts for input and output tokens of every model invocation. Native tool calls without direct model-token consumption do not debit the token ledger; their invocation counts, durations, and payload sizes are independently constrained. Tool outputs incorporated into subsequent model requests are charged as input tokens. The Gateway and trusted infrastructure also enforce maximum output-token allowances, request timeouts, tool-result size limits, invocation counts, and global deadlines, preventing zero-token tool operations from becoming unbounded resource channels.

\subsubsection{Four-Stage Call Protocol}

Each invocation follows four stages under the principle: \emph{reserve before spend, settle after execution, and deliver only after verification}. Reservations and ledger transitions are atomic to prevent concurrent overspending.

\textit{(i)} \textbf{Reservation:} The Gateway computes the complete request quote and reserves token credit from the active ordinary node account or recovery pool. Native tools without direct token consumption reserve the applicable execution allowance and dispatch slot. The reservation and call intent are durably recorded before dispatch.

\textit{(ii)} \textbf{Dispatch:} The Gateway persists the dispatch state before invoking the model or tool. If the outcome becomes unknown after dispatch, the reservation remains pending until execution and resource liabilities are reconciled, preventing blind re-dispatch.

\textit{(iii)} \textbf{Settlement:} Upon receiving a verifiable execution receipt, the Gateway records actual input/output token consumption regardless of application-level success or failure. Settled consumption moves to the global expenditure account, unused reservations return to their originating accounts, and unconfirmed costs remain pending liabilities.

\textit{(iv)} \textbf{Verification and Release:} The Gateway verifies result-delivery eligibility and relevant post-execution side effects before releasing outputs. Unverified or isolated results remain withheld even when their computational costs have been settled.

\subsubsection{Global Token Conservation Model}

The global ledger accounts exclusively for model-token credits. Let $H$ denote the global token cap, $S_{\mathrm{settled}}$ settled input/output consumption across all model invocations, and $P_{\mathrm{inflight}}$ outstanding reservations and pending execution liabilities.

We distinguish three categories of unspent credits:

\begin{itemize}
    \item $A_{\mathrm{node}}$: Unspent credits held in authorized ordinary node accounts.
    \item $A_{\mathrm{ordinary}}$: Unallocated credits in the ordinary resource pool.
    \item $A_{\mathrm{recovery}}$: Unspent credits in the shared recovery pool.
\end{itemize}

The global ledger satisfies

\begin{equation}
H=S_{\mathrm{settled}}+P_{\mathrm{inflight}}+A_{\mathrm{node}}+A_{\mathrm{ordinary}}+A_{\mathrm{recovery}}.
\label{eq:conservation}
\end{equation}

Upon one-time authorization $B_v^{\mathrm{req}}$, credits move atomically from $A_{\mathrm{ordinary}}$ to $A_{\mathrm{node}}$. Before each model invocation, its complete quote moves to $P_{\mathrm{inflight}}$. After verifying the outcome and token usage, actual consumption moves to $S_{\mathrm{settled}}$, while unused reservations return to their originating accounts. Unused ordinary node credits return to $A_{\mathrm{ordinary}}$ after formal settlement.

Recovery calls follow the same protocol but draw exclusively from $A_{\mathrm{recovery}}$; settled consumption enters $S_{\mathrm{settled}}$, while unused reservations return only to the recovery pool. Recovery cannot replenish its initial allotment or transfer remaining credits into the ordinary pool.

In-flight reservations and unresolved liabilities remain distinct from settled consumption, preventing double-counting and premature resource release. Unconfirmed costs or unresolved side effects remain pending and block subsequent dispatches that could compromise resource or execution state. Concurrent reservations preserve the invariant through atomic ledger updates.

Under the hard request-quote contract in Section~\ref{sec:budget}, actual consumption cannot exceed the reserved quote. Therefore, for runs satisfying the trusted execution contract,

\begin{equation}
C_{\mathrm{total}}=S_{\mathrm{settled}}\leq H.
\label{eq:global-token-bound}
\end{equation}

If a receipt reports an excess $\Delta_{\mathrm{breach}}=c_a-q_a>0$, the Gateway records an infrastructure-contract violation, freezes further dispatch, and terminates the affected run. The normal global-bound guarantee does not apply to that violating run.

Non-token constraints, including invocation counts, timeouts, tool-result sizes, and global deadlines, are maintained through separate runtime counters and policies, not included in Equation~\ref{eq:conservation}.

\subsubsection{Anomaly Detection and Control Isolation}

The Gateway performs deterministic, low-overhead anomaly monitoring without external model evaluators. By auditing normalized call arguments, execution outputs, and verified state changes over rolling windows, the Host detects zero-progress loops, repeated deterministic errors, and identical calls when pre-registered thresholds are reached.

Upon detecting a qualified anomaly or state corruption, the Host freezes ordinary scheduling, isolates affected results, and reconciles pending resource liabilities and external side effects. Separating tool execution, result delivery, and subsequent model invocation prevents unverified results from triggering further execution while state and resource records remain unresolved.

Configuration gaps, such as mismatched tool schemas, unsupported model identities, or unverified execution configurations, trigger immediate termination rather than recovery under an untrusted setup. Transient failures may enter recovery only when independently observed, relevant outcomes reconciled, and trusted safe-retry conditions in Section~\ref{sec:recovery} satisfied.

\subsection{Bounded Recovery}
\label{sec:recovery}

When budget exhaustion, a detected anomaly, or an eligible transient failure interrupts a node, the system chooses between termination and controlled recovery. Recovery cannot draw from the ordinary pool, as adversaries could otherwise trigger interruptions to obtain fresh ordinary resources. Resetting the global ledger, execution history, or call counters would create similar exhaustion opportunities.

We therefore introduce a shared, non-resettable recovery pool for all interrupted nodes and attempts within a run. Retries, successful attempts, restarts, and identity transformations cannot replenish it. Recovery resumes interrupted nodes within the original task, preserving verified progress rather than restarting the task. It is best-effort: neither node completion nor end-to-end success is guaranteed.

\subsubsection{Recovery Pool Formulation}

At run initialization, the Host locks the recovery allotment and attempt bound into the run identity. Given global token cap $H$ and pre-registered recovery fraction $\rho\in(0,1)$,

\begin{equation}
R_0=\lfloor\rho H\rfloor,\qquad E_0=H-R_0,
\label{eq:recovery-pool}
\end{equation}

with run-level attempt bound

\begin{equation}
N_{\mathrm{recovery}}\leq K,
\end{equation}

where $K$ applies across the entire run, not per node. All recovery attempts share the same token pool, attempt counter, and execution deadline.

The recovery pool satisfies

\begin{equation}
R_0=S_R+P_R+A_R,
\label{eq:recovery-conservation}
\end{equation}

where $S_R$ is settled recovery consumption, $P_R$ outstanding recovery reservations and pending token liabilities, and $A_R$ remaining recovery credit, corresponding to $A_{\mathrm{recovery}}$ in the global ledger. Unresolved external side effects are tracked separately.

Recovery calls follow the ordinary reservation and settlement protocol but draw exclusively from $A_{\mathrm{recovery}}$. Unused reservations return only to this pool and cannot replenish its initial allotment. Each attempt atomically consumes one opportunity regardless of outcome, and only one attempt may be active at a time.

\subsubsection{Strict Admission Control}

Recovery admission requires strict preconditions to prevent operational failures from becoming repeated resource-consuming execution cycles.

An attempt may be considered following one of three independently observable conditions:

\textit{(i)} \textbf{Budget Exhaustion:} The interrupted node has exhausted its ordinary token budget.

\textit{(ii)} \textbf{Detected Execution Anomaly:} The Gateway or Host detects an anomaly meeting pre-registered interruption thresholds, such as persistent zero-progress loops or repeated deterministic failures.

\textit{(iii)} \textbf{Eligible Transient Operational Failure:} An independently observed transient infrastructure or tool failure, such as a timeout or temporary service error, satisfies configured safe-retry conditions.

Regardless of the trigger, admission requires independently observed failure evidence, frozen ordinary scheduling, and reconciliation of outstanding token liabilities. External side effects are separately classified against trusted business-state observations. Confirmed unexecuted operations may be retried under original permissions; successful operations are preserved and not repeated. Unknown outcomes require trusted state verification before retry; if an outcome cannot be established, blind re-execution is prohibited.

The Host also re-verifies the target node, prerequisites, and original user permissions. Recovery remains subject to the run-level attempt bound $K$, remaining shared recovery credits, and original global deadline. Repeated deterministic failures require new trusted evidence or independently verified safe-retry conditions. Node renaming and agent identity changes cannot bypass the shared attempt counter. A recovery ticket defines a bounded execution envelope, not blanket authorization: every invocation must satisfy original permissions and pass through Gateway reservation and settlement.

\subsubsection{Trusted Reconstruction and Termination Semantics}

The Host reconstructs a fresh execution context from a verified state anchor, delivered public evidence, and confirmed operation outcomes, excluding withheld or isolated content. Private states may veto conflicting facts but are never injected into the model context. Reconstruction relies on trusted business-state observations and execution records, while preserving the original authorization DAG, tool permissions, and task objectives.

\subsubsection{Recovery Settlement and Termination}

After recovery, the Host verifies the business state, execution receipts, and side effects against the original node completion predicate. It issues a new Host-authenticated completion certificate only if that predicate is fully satisfied. Otherwise, the node remains incomplete and cannot unlock successors; further attempts must independently satisfy admission and resource constraints.

Successful recovery does not grant a new ordinary budget. Settled recovery consumption remains charged to the shared recovery pool, while unused ordinary node credits follow the original settlement rules and are never reissued to the interrupted node.

Recovery terminates when the pool cannot cover another request, the attempt limit $K$ is reached, the global deadline expires, or no trusted evidence supports further attempts. None of these conditions resets the ordinary budget, recovery pool, global ledger, or execution deadline.

\section{Experiments}
\label{sec:experiments}

\begin{table*}[t]
\centering
\caption{Task success rates and token consumption under benign tasks. Token counts are reported in thousands. DAG denotes the baseline agent augmented with the public task DAG, while BA denotes the proposed Bounded Autonomy framework.}
\label{tab:benign-results}

\begin{tabular}{cc*{12}{c}}
\toprule
\multirow{3}{*}{Model}
& \multirow{3}{*}{Agent}
& \multicolumn{2}{c}{Telecom}
& \multicolumn{2}{c}{Airline}
& \multicolumn{2}{c}{Retail}
& \multicolumn{2}{c}{Banking Knowledge}
& \multicolumn{2}{c}{Mock}
& \multicolumn{2}{c}{All} \\
\cmidrule(lr){3-4}
\cmidrule(lr){5-6}
\cmidrule(lr){7-8}
\cmidrule(lr){9-10}
\cmidrule(lr){11-12}
\cmidrule(lr){13-14}

& & \multirow{2}{*}{TSR} & Tokens
& \multirow{2}{*}{TSR} & Tokens
& \multirow{2}{*}{TSR} & Tokens
& \multirow{2}{*}{TSR} & Tokens
& \multirow{2}{*}{TSR} & Tokens
& \multirow{2}{*}{TSR} & Tokens \\

& &
& ($10^3$)
& & ($10^3$)
& & ($10^3$)
& & ($10^3$)
& & ($10^3$)
& & ($10^3$) \\
\midrule

\multirow{3}{*}{Qwen}
& DAG
& 0.87 & 277.0
& 0.90 & 94.5
& 0.69 & 92.8
& 0.45 & 234.5
& 1.00 & 6.1
& 0.76 & 179.4 \\

& BA
& 0.84 & 302.2
& 1.00 & 97.0
& 0.64 & 90.7
& 0.50 & 187.6
& 1.00 & 3.4
& 0.76 & 179.8 \\

& \textit{Change}
& $-0.03$ & $\uparrow 9.1\%$
& $+0.10$ & $\uparrow 2.6\%$
& $-0.05$ & $\downarrow 2.3\%$
& $+0.05$ & $\downarrow 20.0\%$
& $0.00$ & $\downarrow 44.3\%$
& $0.00$ & $\uparrow 0.2\%$ \\

\midrule

\multirow{3}{*}{DeepSeek}
& DAG
& 0.89 & 177.1
& 1.00 & 112.2
& 0.85 & 72.3
& 0.45 & 232.1
& 1.00 & 5.5
& 0.83 & 139.3 \\

& BA
& 0.91 & 185.2
& 0.95 & 98.3
& 0.92 & 70.4
& 0.50 & 219.6
& 1.00 & 2.8
& 0.86 & 137.4 \\

& \textit{Change}
& $+0.02$ & $\uparrow 4.6\%$
& $-0.05$ & $\downarrow 12.4\%$
& $+0.07$ & $\downarrow 2.6\%$
& $+0.05$ & $\downarrow 5.4\%$
& $0.00$ & $\downarrow 49.1\%$
& $+0.03$ & $\downarrow 1.4\%$ \\

\bottomrule
\end{tabular}
\end{table*}

\subsection{Experimental Setup}

\subsubsection{Datasets}

We use $\tau^2$-bench~\cite{26dddae50d5a4610bdaced3e9fdb1c89} v1.0.1 as our evaluation benchmark, which assesses the ability of LLM agents to accomplish tasks through tool interactions in interactive environments. Our experiments cover five domains: \texttt{airline}, \texttt{retail}, \texttt{telecom}, \texttt{banking\_knowledge}, and \texttt{mock}. Tasks in each domain are partitioned into training, validation, and test sets. The training set is used to collect benign execution trajectories and construct the corresponding reference budget estimator, while the validation set is used to select and freeze experimental configurations. The test set is reserved exclusively for final evaluation.

For budget estimation, each individual task in a domain's training split constitutes one statistical observation unit, with all execution nodes belonging to the same task retained together during fitting and calibration. Budget estimation is performed independently for each domain. Task success is determined by the official $\tau^2$-bench evaluator. Detailed dataset statistics and partitioning information are provided in the Appendix.

\subsubsection{Baselines}

We compare two agent execution configurations: \textsc{DAG-Agent} and \textsc{BA-Agent} (Bounded Autonomy). DAG-Agent serves as a structure-matched baseline to control for the effects of explicit DAG-based task organization, allowing us to isolate the additional benefits of the proposed security and resource-governance mechanisms.

\textbf{\textsc{DAG-Agent}.} This baseline builds on the native $\tau^2$-bench agent and introduces only the pre-constructed public task DAG. The agent executes tasks according to the exposed node objectives, dependencies, and allowed tool sets, retaining the benchmark's native tool permissions, sandboxing, and error handling, but without the proposed Host-authenticated progress certificates, one-time node budgets, Gateway-based resource enforcement, or bounded recovery.

\textbf{\textsc{BA-Agent}.} This configuration enables the complete Bounded Autonomy defense framework proposed in this paper.

Both configurations use identical task instances, initial environment states, pre-constructed public DAGs, model backbones, tool interfaces, user simulators, random seeds, decoding parameters, maximum execution steps, and outer execution deadlines. They also share the same task contracts and tool permissions, ensuring a paired comparison under otherwise matched execution conditions.

\subsubsection{Attack Settings}

We evaluate three attacks independently, each with multiple intensity levels. Each attack is paired with a clean condition without adversarial injection, and identical attack configurations are applied to both DAG-Agent and BA-Agent. The attacks are simulated by injecting adversarial content or error feedback through normal tool interactions.

\textit{(i) Persistent Context Amplification:} Each injected payload has a fixed size of 128 KiB, and all agent-initiated tool calls are eligible for injection. We consider four intensity levels, corresponding to a maximum of 1, 2, 3, and 4 injections per task.

\textit{(ii) Large Tool Output Amplification:} Repetitive text is injected into schema-valid tool responses, with at most one injection per task. We consider four payload sizes: 64 KiB, 128 KiB, 256 KiB, and 512 KiB.

\textit{(iii) Error-Induced Agentic Loops:} Timeout errors are synthetically injected into agent-initiated tool calls. We consider four intensity levels, corresponding to 1, 2, 3, and 4 injected errors per task.

\subsubsection{Agent Settings}

We use Qwen3.8-Flash~\cite{qiu2026design} and DeepSeek-V4.1-Flash~\cite{xu2026deepseek} as the agent backbones. For each combination of model backbone and evaluation domain, we independently collect benign training trajectories and construct a corresponding reference budget estimator. Each estimator is fitted and calibrated using the corresponding domain's training tasks, with its hyperparameters and runtime configurations selected on the respective validation set and frozen before test evaluation. Neither training data nor fitted estimators are shared across model backbones or evaluation domains. Detailed model configurations and execution parameters are provided in the Appendix.

\subsubsection{Evaluation Metrics}

We evaluate the effectiveness of our defense framework using four metrics: task success rate, token consumption, and two recovery-related metrics.

\textbf{Task Success Rate.} We use the official $\tau^2$-bench evaluator to determine task completion. A task is considered successful if its final reward equals 1, and unsuccessful otherwise. The task success rate is defined as

\begin{equation}
\mathrm{TSR}=\frac{N_{\mathrm{success}}}{N_{\mathrm{total}}},
\end{equation}

where $N_{\mathrm{success}}$ and $N_{\mathrm{total}}$ denote the number of successfully completed tasks and the total number of evaluated tasks, respectively. Node-level completion and successful recovery alone do not constitute task success.

\textbf{Token Consumption.} We measure token consumption using actual usage reported by model providers, including both input and output tokens across all model invocations. This includes normal execution, auxiliary control calls, and recovery-related model calls, including actual token consumption reported for unsuccessful invocations. Authorized token credits are distinguished from realized consumption, and unused reservations are not counted as consumed tokens. Tokens used exclusively for offline estimator construction and task evaluation are excluded.

\textbf{Recovery Metrics.} We evaluate recovery performance using two metrics. Node Recovery Success Rate (NRSR) measures the proportion of recovery-admitted logical nodes that eventually satisfy their original Host-defined completion predicates. Each logical node is counted only once, regardless of the number of recovery attempts associated with it. Recovery Pool Utilization (RPU) measures the aggregate fraction of the initially allocated recovery credits consumed by actual recovery-related model invocations across the evaluated runs. These metrics are defined as

\begin{equation}
\begin{aligned}
\mathrm{NRSR}
&=
\frac{N_{\mathrm{recovered\ nodes}}}
{N_{\mathrm{admitted\ nodes}}}
\times 100\%,\\[6pt]
\mathrm{RPU}
&=
\frac{
\sum_{r\in\mathcal R} C_{\mathrm{recovery}}^{(r)}
}{
\sum_{r\in\mathcal R} R_0^{(r)}
}
\times 100\%.
\end{aligned}
\end{equation}

Here, $N_{\mathrm{recovered\ nodes}}$ denotes the number of distinct logical nodes that eventually satisfy their original Host-defined completion predicates, and $N_{\mathrm{admitted\ nodes}}$ denotes the number of distinct logical nodes admitted into recovery. $\mathcal R$ denotes the set of evaluated experimental runs, $C_{\mathrm{recovery}}^{(r)}$ denotes the actual settled token consumption of recovery-related model invocations in run $r$, and $R_0^{(r)}$ denotes the initial recovery allocation for that run. The RPU is calculated by aggregating recovery consumption and initial recovery allocations across all evaluated runs within each attack category, including all attack intensity levels.

\begin{figure*}[!t]
    \centering
    \includegraphics[width=1.0\linewidth]{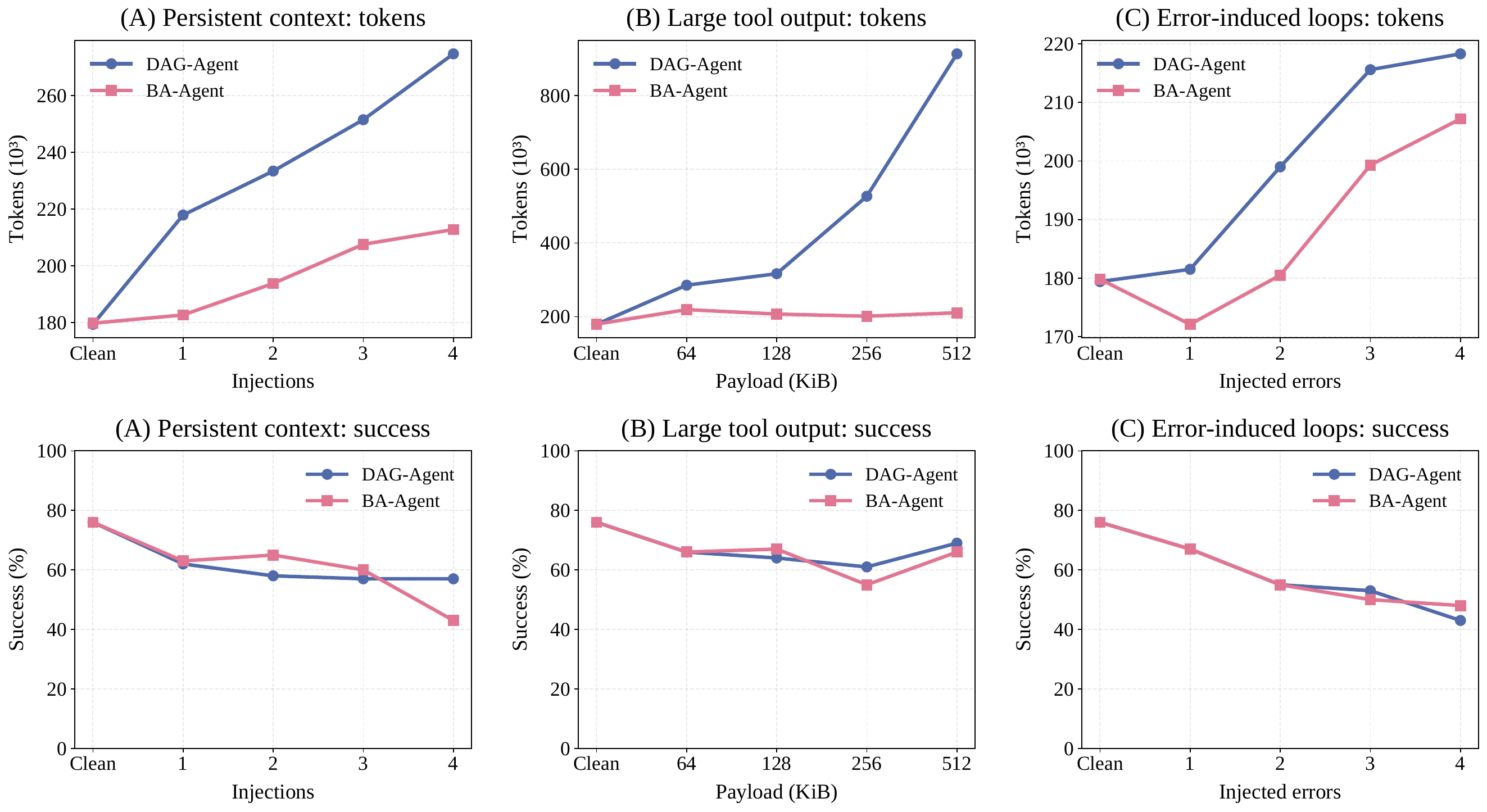}
    \caption{Token consumption (top row) and task success rates (bottom row) of DAG-Agent and BA-Agent under three resource-exhaustion attacks with varying intensities using Qwen. From left to right: Persistent Context Amplification, Large Tool Output Amplification, and Error-Induced Agentic Loops.}
    \label{fig:3}
\end{figure*}

\subsection{Evaluation on Benign Tasks}
\label{sec:benign-evaluation}

We first evaluate the performance of BA-Agent under benign conditions, where no adversarial payloads are injected. Table~\ref{tab:benign-results} presents the task success rates and token consumption of both agents across five domains and two model backbones.

\textbf{Task Success Rate.} Overall, BA-Agent maintains comparable task success rates to DAG-Agent across both model backbones. With Qwen, both agents achieve an overall success rate of 0.76. With DeepSeek, BA-Agent achieves 0.86, compared with 0.83 for DAG-Agent. At the domain level, the results vary across models. For Qwen, BA-Agent improves task success rates in Airline and Banking Knowledge by 10 and 5 percentage points, respectively, while experiencing decreases of 3 and 5 percentage points in Telecom and Retail. For DeepSeek, BA-Agent improves performance in Telecom, Retail, and Banking Knowledge by 2, 7, and 5 percentage points, respectively, while Airline experiences a 5-percentage-point decrease. Both agents achieve a success rate of 1.00 in Mock. These results indicate that the proposed defense preserves overall task utility, although its impact on individual domains varies with the underlying model and task characteristics.

\textbf{Token Consumption.} In terms of resource consumption, BA-Agent exhibits only minor changes in overall token usage compared with DAG-Agent. With Qwen, token consumption increases by only 0.2\%, while with DeepSeek, it decreases by 1.4\%. At the domain level, BA-Agent reduces token consumption in most domains. In particular, Mock exhibits the largest reductions, reaching 44.3\% and 49.1\% under Qwen and DeepSeek, respectively. Banking Knowledge also exhibits notable reductions of 20.0\% and 5.4\%, respectively. In contrast, Telecom incurs higher token consumption under both backbones, increasing by 9.1\% and 4.6\%, respectively. These variations suggest that the resource overhead introduced by the defense is workload-dependent. Nevertheless, the overall results demonstrate that BA-Agent can preserve benign task utility without introducing substantial additional token consumption.

\subsection{Evaluation under Attacks}
\label{sec:attack-evaluation}

We evaluate BA-Agent against three resource-exhaustion attacks under different intensity levels using both Qwen and DeepSeek. For clarity of presentation, we report the Qwen results in the main text, while the corresponding results and analyses for DeepSeek are provided in the Appendix. Figure~\ref{fig:3} presents token consumption and task success rates under Persistent Context Amplification, Large Tool Output Amplification, and Error-Induced Agentic Loops. Overall, BA-Agent consistently limits attack-induced token consumption, although the magnitude of resource reduction and its impact on task utility vary across attack types.

\textbf{Persistent Context Amplification.}
As the number of injected payloads increases, DAG-Agent exhibits a steady increase in token consumption, from 179.4K under the clean condition to 274.7K at the highest intensity. In contrast, BA-Agent limits consumption to 212.8K, reducing token usage by 22.5\% relative to DAG-Agent at the highest intensity. This result indicates that the proposed framework mitigates the cumulative resource overhead caused by persistent adversarial content. In terms of task success, BA-Agent achieves comparable or slightly higher rates than DAG-Agent at the first three attack intensities. However, at the highest intensity, its success rate drops to 43\%, compared with 57\% for DAG-Agent. This suggests that under severe persistent context amplification, stronger resource containment may come at the cost of task completion utility.

\textbf{Large Tool Output Amplification.}
BA-Agent achieves its most substantial resource reduction under large tool output amplification. As the injected payload size increases from 64 KiB to 512 KiB, DAG-Agent's token consumption rises sharply from 285.5K to 912.8K. In contrast, BA-Agent maintains consumption within a relatively narrow range of 201.4K--219.3K. At the largest payload size, BA-Agent reduces token consumption by 76.9\% compared with DAG-Agent. Meanwhile, task success rates remain broadly comparable between the two configurations: BA-Agent matches DAG-Agent at the first intensity and achieves a slightly higher rate at the second, while showing decreases of 6 and 3 percentage points at the third and fourth intensities, respectively. These results demonstrate that BA-Agent substantially limits resource amplification caused by oversized tool outputs while largely preserving task utility.

\textbf{Error-Induced Agentic Loops.}
Under error-induced agentic loops, the token consumption gap between the two configurations is smaller than that observed under the other attacks. At the highest intensity, DAG-Agent consumes 218.3K tokens, whereas BA-Agent consumes 207.2K, corresponding to a reduction of 5.1\%. Nevertheless, BA-Agent consistently incurs lower token consumption across all non-zero error intensities. Regarding task success, the two configurations achieve identical rates at the first two intensities. At the third intensity, BA-Agent is 3 percentage points lower, whereas at the highest intensity it achieves 48\%, compared with 43\% for DAG-Agent. This indicates that the proposed framework provides moderate resource containment under error-induced loops and can maintain task utility under severe error feedback.

Overall, the results show that the effectiveness of BA-Agent depends on the attack mechanism. Its strongest resource-containment effect occurs under large tool output amplification, where it prevents the substantial token growth observed in DAG-Agent. It also reduces resource consumption under persistent context amplification and error-induced loops, although the associated task utility varies with attack intensity. These findings demonstrate that the proposed framework provides resource control across different attack mechanisms, while highlighting the trade-off between resource containment and task completion under severe adversarial conditions.

\subsection{Evaluation of Bounded Recovery}
\label{sec:recovery-evaluation}

We evaluate the bounded recovery mechanism under the three resource-exhaustion attacks using Qwen. Results are aggregated across all intensity levels within each attack category. We examine node recovery effectiveness, recovery resource consumption, and the impact of recovery on end-to-end task success.

\textbf{Recovery Effectiveness.} Table~\ref{tab:recovery-results} summarizes the recovery performance of BA-Agent. The Node Recovery Success Rate (NRSR) reaches 77\%, 81\%, and 59\% under Persistent Context Amplification, Large Tool Output Amplification, and Error-Induced Agentic Loops, respectively. These results indicate that most nodes admitted to recovery eventually satisfy their original Host-defined completion predicates. Recovery performs particularly well under large tool output amplification, achieving the highest NRSR of 81\%.

Error-induced agentic loops present greater recovery challenges, with an NRSR of 59\%. This attack induces repeated tool failures, redundant reads, and execution sequences that fail to produce verifiable business-state changes. Although these nodes are admitted into recovery, the Gateway continues to monitor their execution and terminates recovery when repeated failures or the absence of meaningful progress persist. Consequently, some recovery-admitted nodes fail to satisfy their original completion predicates.

\begin{table}[t]
\centering
\small
\caption{Recovery performance of BA-Agent under three resource-exhaustion attacks using Qwen. Results are aggregated across all intensity levels.}
\label{tab:recovery-results}
\begin{tabular}{lcc}
\toprule
\textbf{Attack} & \textbf{NRSR (\%)} & \textbf{RPU (\%)} \\
\midrule
Persistent Context Amplification & 77 & 39 \\
Large Tool Output Amplification & 81 & 19 \\
Error-Induced Agentic Loops & 59 & 32 \\
\bottomrule
\end{tabular}
\end{table}

\textbf{Recovery Resource Consumption.} The corresponding RPU values are 39\%, 19\%, and 32\%, respectively. The relatively low utilization under large tool output amplification is primarily attributable to its single-shot attack pattern, which generally requires only one recovery operation. In contrast, persistent context amplification and error-induced loops can trigger repeated interruptions at higher attack intensities, resulting in additional recovery attempts and higher resource utilization. Importantly, the recovery pool represents an upper resource bound rather than a consumption target. All three attack categories consume less than 40\% of their initial recovery allocations, indicating that recovery remains within its dedicated resource budget.

\textbf{Impact on Task Success.} To assess the contribution of recovery to end-to-end task completion, we compare BA-Agent with and without the recovery mechanism. Table~\ref{tab:recovery-ablation} reports the task success rates aggregated across attack intensity levels. Enabling recovery improves task success by 37, 41, and 24 percentage points under Persistent Context Amplification, Large Tool Output Amplification, and Error-Induced Agentic Loops, respectively. The improvement is observed across all three attacks, with the largest gain under large tool output amplification. These results demonstrate that bounded recovery helps interrupted executions resume and complete their original tasks, rather than merely providing additional execution attempts.

\begin{table}[t]
\centering
\small
\caption{Task success rates of BA-Agent with and without bounded recovery under three attacks using Qwen. Results are aggregated across all intensity levels.}
\label{tab:recovery-ablation}
\begin{tabular}{lccc}
\toprule
\textbf{Configuration} & \textbf{A (\%)} & \textbf{B (\%)} & \textbf{C (\%)} \\
\midrule
With Recovery & 58 & 64 & 55 \\
Without Recovery & 21 & 23 & 31 \\
\midrule
Improvement & +37 & +41 & +24 \\
\bottomrule
\end{tabular}
\end{table}

Overall, these results show that bounded recovery improves task completion under all three attacks while maintaining limited recovery resource consumption. Combined with the node-level recovery results, they demonstrate that the shared recovery pool provides meaningful recovery opportunities without introducing unbounded resource renewal.

\section{Conclusion}

This paper presented BA-Agent (Bounded Autonomy), a trust-anchored defense framework for mitigating resource-exhaustion attacks against tool-augmented LLM agents. BA-Agent separates agent autonomy from system-level resource governance through four complementary mechanisms: Host-compiled authorization DAGs and authenticated completion certificates for verified task progression, calibrated one-time node budgets for upfront resource allocation, centralized Gateway enforcement for deterministic resource accounting, and a shared, non-resettable recovery pool for bounded exception handling. These mechanisms jointly establish resource boundaries while preserving the agent's ability to autonomously plan and execute tasks.

Our evaluation demonstrates that BA-Agent can effectively limit adversarial resource consumption while maintaining comparable utility under benign conditions. By combining upfront resource authorization, runtime enforcement, and controlled recovery, BA-Agent addresses resource exhaustion as a system-level execution governance problem rather than relying solely on the agent's behavioral robustness. Although recovery remains best-effort under severe adversarial conditions, the framework provides a practical foundation for building more resource-bounded and resilient tool-augmented LLM agent systems.



%
\bibliographystyle{IEEEtran}
\bibliography{references}

\appendices

\section{Budget Estimator Hyperparameters}

\label{app:budget-hyperparameters}

Table~\ref{tab:budget-hyperparameters} summarizes the
hyperparameter configuration of the reference demand
estimator. We adopt a unified set of hyperparameters
across different model backbones and evaluation domains.
Specifically, the reference quantile level is set to
$\tau=0.95$, the residual normalization scale to
$s_0=1$, the node-budget safety margin to
$\epsilon=0.02$, and the upper-tail miscoverage level
to $\alpha=0.05$. We use $J=5$ folds for cross-fitting.

Note that $\tau$ and $\alpha$ govern different stages
of the estimation procedure: $\tau$ determines the
reference demand quantile, whereas $\alpha$ controls
the upper-tail calibration level.

For each model backbone and evaluation domain, the
reference demand estimator is independently fitted
and calibrated using benign execution trajectories
from the corresponding training set. The hyperparameter
values are fixed before test evaluation.

\section{Runtime Resource Configuration}
\label{app:runtime-configuration}

Table~\ref{tab:runtime-configuration} summarizes the
primary runtime resource limits adopted in our
experiments. The recovery pool is allocated 10\% of
the global token budget, with at most three recovery
attempts allowed per task run. To accommodate different
model output capacities, the maximum output-token
allowance is set to 16K for Qwen and 32K for DeepSeek.

In addition, the Gateway enforces a maximum of 100
model invocations and 400 tool invocations per task run.
The maximum number of execution steps is also limited
to 100 per task run. The outer process deadline is set
to 2,100 seconds per task run, while each individual
tool execution is subject to a timeout of 120 seconds.

Other model- or domain-specific runtime configurations,
including the global token-budget cap $H$, are
independently determined using the corresponding
validation sets and frozen before test evaluation.
All configurations remain fixed throughout the final
test evaluation, without being adjusted based on
test outcomes.

\begin{table}[t]
\centering
\caption{Hyperparameter Configuration of the Reference
Demand Estimator.}
\label{tab:budget-hyperparameters}
\small
\begin{tabular}{c c c}
\hline
\textbf{Parameter} & \textbf{Description} & \textbf{Value} \\
\hline
$\tau$ & Reference quantile level & 0.95 \\
$s_0$ & Residual normalization scale & 1 \\
$\epsilon$ & Node-budget safety margin & 0.02 \\
$\alpha$ & Upper-tail miscoverage level & 0.05 \\
$J$ & Number of cross-fitting folds & 5 \\
\hline
\end{tabular}
\end{table}

\begin{table}[t]
\centering
\caption{Runtime Resource Configuration.}
\label{tab:runtime-configuration}
\begin{tabular}{
    >{\centering\arraybackslash}p{0.22\columnwidth}
    >{\centering\arraybackslash}p{0.49\columnwidth}
    >{\centering\arraybackslash}p{0.17\columnwidth}
}
\hline
\textbf{Parameter} & \textbf{Description} & \textbf{Value} \\
\hline
$\rho$ & Recovery pool fraction & 10\% \\
$K$ & Maximum recovery attempts per run & 3 \\
$O_{\max}$ & Maximum output tokens (Qwen / DeepSeek) & 16K / 32K \\
$N_{\mathrm{model}}^{\max}$ & Maximum model invocations per run & 100 \\
$N_{\mathrm{tool}}^{\max}$ & Maximum tool invocations per run & 400 \\
$N_{\mathrm{step}}^{\max}$ & Maximum execution steps per run & 100 \\
$T_{\mathrm{run}}^{\max}$ & Outer process deadline per run & 2100 s \\
$T_{\mathrm{tool}}^{\max}$ & Tool execution timeout per invocation & 120 s \\
\hline
\end{tabular}
\end{table}

\section{Attack Configuration}
\label{app:attack-configuration}

This appendix describes the construction and injection
procedures of the three resource-exhaustion attacks
evaluated in our experiments. All attacks are delivered
through the agent's normal interaction interfaces,
without modifying the underlying task definitions or
benchmark evaluation logic.

\textbf{Persistent Context Amplification.}
This attack induces cumulative context growth by
injecting adversarial messages into subsequent model
requests following agent-initiated tool calls. The
original tool responses are preserved unchanged, while
an additional message with role \texttt{user} and name
\texttt{abc\_synthetic\_context} is appended to the
next model request. Each payload contains instructions
encouraging the agent to retain complete records across
subsequent turns, together with repetitive record-
comparison content. The payload is padded to exactly
128~KiB. Its content and size remain identical across
tasks, model backbones, and execution configurations.
We vary the attack intensity exclusively by changing
the maximum number of injections per task.

\textbf{Large Tool Output Amplification.}
This attack increases the volume of tool-returned
content by injecting repetitive text into pre-audited
and registered textual fields of tool responses. The
payload is constructed by repeating the fixed string
``Record retained for redundancy verification.''
and padding it to the specified payload size. For
structured JSON responses, the injected content is
appended within the designated string field while
preserving the remaining fields and the response
schema. For plain-text responses, the content is
appended directly to the original response. We
evaluate four payload sizes: 64, 128, 256, and
512~KiB, with at most one injection per task.

\textbf{Error-Induced Agentic Loops.}
This attack induces repeated agent execution by
intercepting tool invocations and injecting synthetic
timeout errors. All agent-initiated tool calls,
including both read and write operations, are within
the attack scope. The interception occurs before the
invocation reaches the underlying tool, causing the
actual tool execution to be skipped. The agent instead
receives a synthetic tool response with
\texttt{error=true}, containing a \texttt{TimeoutError}
indication and an instruction to retry the operation.
Each injected error message is padded to 256 bytes.
The synthetic error is generated immediately upon
interception and does not require waiting for an
actual timeout to occur. Since intercepted invocations
are not executed by the underlying tools, they do not
produce native tool-side business effects. We vary
the attack intensity by injecting 1, 2, 3, or 4
synthetic timeout errors per task.

\section{Dataset Statistics and Data Partitioning}
\label{app:dataset-statistics}

Table~\ref{tab:dataset-statistics} summarizes the
dataset statistics and partitioning across the five
evaluation domains. In total, our evaluation comprises
569 tasks, including 354 training tasks, 88 validation
tasks, and 127 test tasks. The five domains exhibit
different dataset sizes.

For each domain, tasks are partitioned into training,
validation, and test sets. The training set is used to
collect benign execution trajectories and fit the
corresponding reference demand estimator. The validation
set is used to select hyperparameters and runtime
configurations, which are subsequently frozen before
test evaluation. The test set is reserved exclusively
for final evaluation and is not used for estimator
fitting, calibration, or configuration selection.

To reduce semantic overlap and potential information
leakage across subsets, task families or underlying
scenarios are grouped during partitioning, ensuring
that related tasks are assigned to the same subset.

\begin{table}[t]
\centering
\caption{Dataset Statistics and Data Partitioning.}
\label{tab:dataset-statistics}
\small
\begin{tabular}{ccccc}
\hline
\textbf{Domain} &
\textbf{Train} &
\textbf{Validation} &
\textbf{Test} &
\textbf{Total} \\
\hline
Airline & 23 & 6 & 20 & 49 \\
Retail & 60 & 15 & 39 & 114 \\
Telecom & 210 & 45 & 45 & 300 \\
Banking Knowledge & 58 & 19 & 20 & 97 \\
Mock & 3 & 3 & 3 & 9 \\
\hline
Total & 354 & 88 & 127 & 569 \\
\hline
\end{tabular}
\end{table}

\begin{figure*}[!t]
    \centering
    \includegraphics[width=1.0\linewidth]{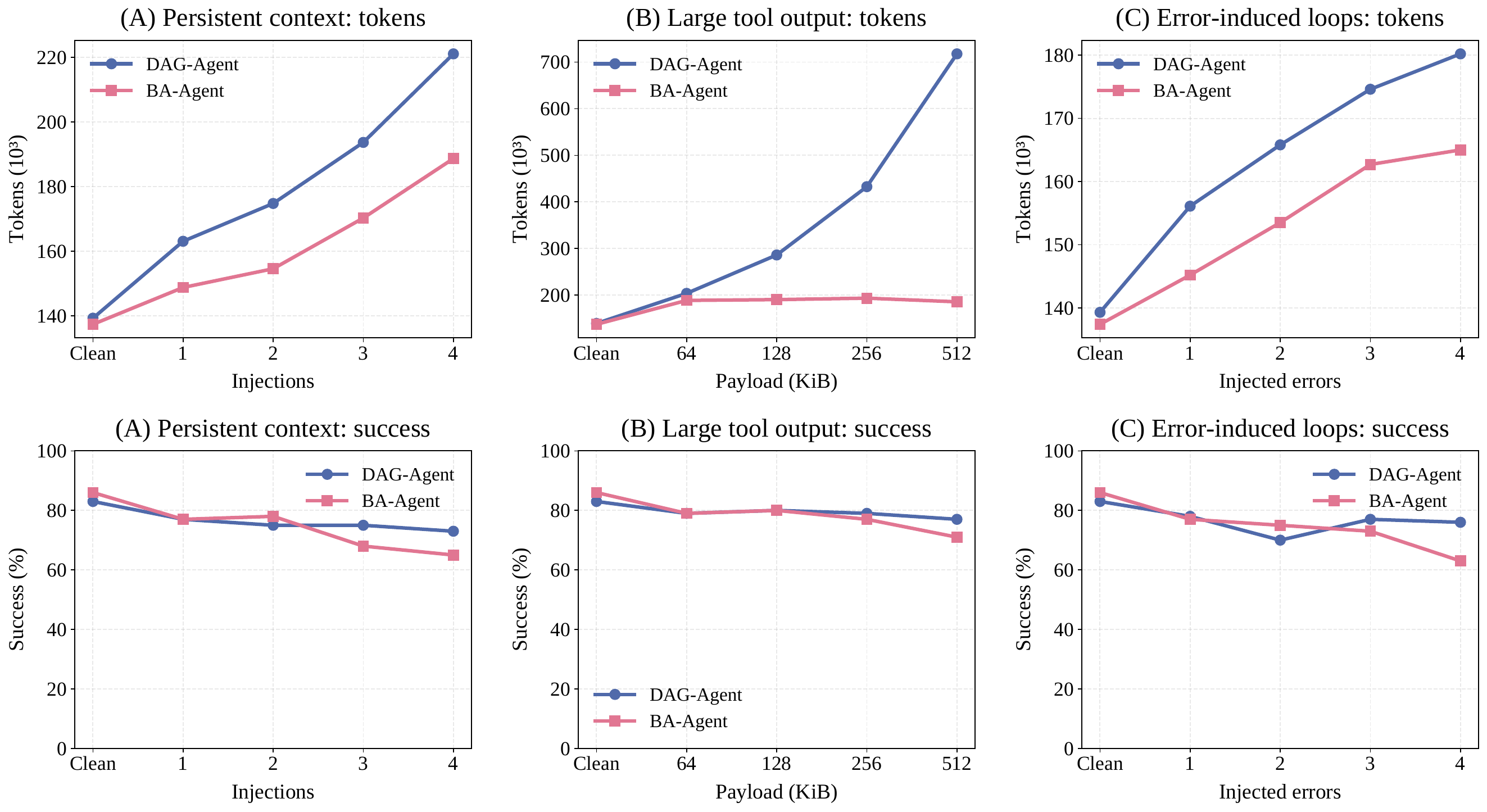}
    \caption{Token consumption (top row) and task success rates (bottom row) of DAG-Agent and BA-Agent under three resource-exhaustion attacks with varying intensities using Deepseek. From left to right: Persistent Context Amplification, Large Tool Output Amplification, and Error-Induced Agentic Loops.}
    \label{fig:4}
\end{figure*}

\section{Additional Attack Evaluation}

To examine whether the proposed framework generalizes across model backbones, we additionally evaluate DAG-Agent and BA-Agent using DeepSeek-V4.1-Flash under the same three resource-exhaustion attacks. Figure~\ref{fig:4} presents token consumption and task success rates across different attack intensities.

\textbf{Resource Consumption.} BA-Agent consistently limits token consumption across all three attacks. The most substantial reduction occurs under Large Tool Output Amplification. At the largest payload size (512 KiB), DAG-Agent consumes approximately 720K tokens, whereas BA-Agent consumes approximately 186K, corresponding to a reduction of about 74\%. Under Persistent Context Amplification, BA-Agent reduces token consumption from approximately 221K to 189K at the highest injection level. A similar but more moderate reduction is observed under Error-Induced Agentic Loops, where consumption decreases from approximately 180K to 165K at the highest intensity. These results confirm that the framework's resource-containment behavior extends to a different model backbone, with the strongest effect against direct tool-output amplification.

\textbf{Task Utility.} Under clean conditions, BA-Agent achieves a task success rate of approximately 87\%, compared with 83\% for DAG-Agent. Under attacks, the success-rate differences vary with attack type and intensity. BA-Agent remains broadly comparable to DAG-Agent under Large Tool Output Amplification at lower and intermediate payload sizes, although a larger gap emerges at the highest intensity. Under Persistent Context Amplification and Error-Induced Agentic Loops, BA-Agent exhibits lower success rates at the highest intensities despite reducing token consumption. These results are consistent with the utility--resource trade-off observed under Qwen and indicate that the degree of task-utility preservation depends on both the attack mechanism and model backbone.

Overall, the DeepSeek results further demonstrate that BA-Agent can constrain attack-induced token growth across different model backbones, particularly under large tool output amplification, while retaining generally comparable task utility under less severe attack conditions.

\end{document}